\documentclass[aip,jcp,amsmath,amssymb,reprint]{revtex4-1}

\usepackage{graphicx}
\usepackage{hyperref}
\usepackage{dcolumn}
\usepackage{labelfig}
\usepackage{here}

\begin{document}

\title[Interplay of intrinsic disorder and crowding on $\alpha$-syn fibril formation]
{The interplay of intrinsic disorder and macromolecular crowding on $\alpha$-synuclein fibril formation}

\author{Nobu C. Shirai}
\email{shirai@cc.mie-u.ac.jp}
\affiliation{Center for Information Technologies and Networks, Mie University, Tsu, Mie 514-8507, Japan}%
\affiliation{Cybermedia Center, Osaka University, Toyonaka, Osaka 560-0043, Japan}%
\affiliation{Department of Biophysics, Division of Biology, Graduate School of Science, Kyoto University, Sakyo, Kyoto 606-8502, Japan}%
\affiliation{Graduate School of Science, Osaka University, Toyonaka, Osaka 560-0043, Japan}%

\author{Macoto Kikuchi}
\affiliation{Cybermedia Center, Osaka University, Toyonaka, Osaka 560-0043, Japan}%
\affiliation{Graduate School of Science, Osaka University, Toyonaka, Osaka 560-0043, Japan}%
\affiliation{Graduate School of Frontier Biosciences, Osaka University, Suita, Osaka 565-0871, Japan}

\date{\today}

\begin{abstract}
$\alpha$-synuclein ($\alpha$-syn) is an intrinsically disordered protein which is considered to be one of the causes of Parkinson's disease. 
This protein forms amyloid fibrils when in a highly concentrated solution. 
The fibril formation of $\alpha$-syn is induced not only by increases in $\alpha$-syn concentration but also by macromolecular crowding. 
In order to investigate the coupled effect of the intrinsic disorder of $\alpha$-syn and macromolecular crowding, we construct a lattice gas model of $\alpha$-syn in contact with a crowding agent reservoir based on statistical mechanics. 
The main assumption is that $\alpha$-syn can be expressed as coarse-grained particles with internal states coupled with effective volume; and disordered states are modeled by larger particles with larger internal entropy than other states. 
Thanks to the simplicity of the model, we can exactly calculate the number of conformations of crowding agents, and this enables us to prove that the original grand canonical ensemble with a crowding agent reservoir is mathematically equivalent to a canonical ensemble without crowding agents. 
In this expression, the effect of macromolecular crowding is absorbed in the internal entropy of disordered states; it is clearly shown that the crowding effect reduces the internal entropy.
Based on Monte Carlo simulation, we provide scenarios of crowding-induced fibril formation. 
We also discuss the recent controversy over the existence of helically folded tetramers of $\alpha$-syn, and suggest that macromolecular crowding is the key to resolving the controversy.
\end{abstract}

\maketitle

\section{Introduction}
$\alpha$-synuclein ($\alpha$-syn) is a 140-amino-acid protein~\cite{UedaSaitoh1993} expressed in presynaptic terminals of the central nervous system~\cite{JakesGoedert1994,IwaiSaitoh1995,KahleHaass2000,LiuYu2009}. 
$\alpha$-syn is known as an intrinsically disordered protein (IDP), which lacks secondary or tertiary structures~\cite{WeinrebLansbury1996,EliezerBrowne2001,UverskyDunker2008,ShiraiKikuchi2013}. 
Growing evidence suggests that $\alpha$-syn has a causative role in Parkinson's disease (PD), which is the second most common neurodegenerative disease after Alzheimer's disease. 
Genetic mutations in $\alpha$-syn (A30P, E46K and A53T) which have been identified in familial PD (FPD) directly link $\alpha$-syn with PD~\cite{KrugerRiess1998,ZarranzYebenes2004,PolymeropoulosNussbaum1997,MunozTolosa1997}. 
Duplication and triplication of the $\alpha$-syn gene are also found in FPD~\cite{Chartier-HarlinDestee2004,IbanezBrice2004,SingletonGwinn-Hardy2003}. 
The intracellular inclusions called Lewy bodies are the pathological hallmark of PD, and they are mainly composed of $\alpha$-syn~\cite{SpillantiniGoedert1997,SpillantiniGoedert1998}. 
Lewy bodies also link $\alpha$-syn with PD. 
Furthermore, {\it in vitro} studies revealed that $\beta$-sheet-rich amyloid fibrils are formed in high-concentration $\alpha$-syn solutions after hours or days of incubation~\cite{HashimotoMasliah1998,RaaijSubramaniam2008}, and that introduction of these fibrils into cultured cells with overexpression of $\alpha$-syn leads to the formation of Lewy body-like inclusions~\cite{LukLee2009}. 
Inhibition of $\alpha$-syn fibril formation is one of the suggested therapeutic approaches for PD~\cite{LashuelMasliah2013,BartelsSelkoe2011,FauvetLashuel2012may,GouldIschiropoulos2014}, and thus a good understanding of such fibril formation is needed. 

Fibril formation of $\alpha$-syn is a nucleation-dependent process~\cite{WoodBiere1999}. 
Nucleation of $\alpha$-syn protofibrils occurs above a certain $\alpha$-syn concentration, which is called the critical concentration, and fibrils grow only after nucleation~\cite{HashimotoMasliah1998,WoodBiere1999,RaaijSubramaniam2008}. 
The fibril formation kinetics are affected by macromolecular crowding, and crowding agents induce and accelerate the fibril formation of $\alpha$-syn~\cite{ShtilermanLansbury2002,UverskyFink2002}. 
White {\it et~al.} observed that fibril formation of IDPs including $\alpha$-syn is accelerated and that of folded proteins is decelerated by macromolecular crowding, and they concluded that accelerated fibril formation is characteristic of IDPs~\cite{WhiteDobson2010}. 
Macromolecular crowding is an important factor in considering the fibril formation of $\alpha$-syn.

\begin{figure}[h]
\centerline{
\includegraphics[width=0.45\textwidth]{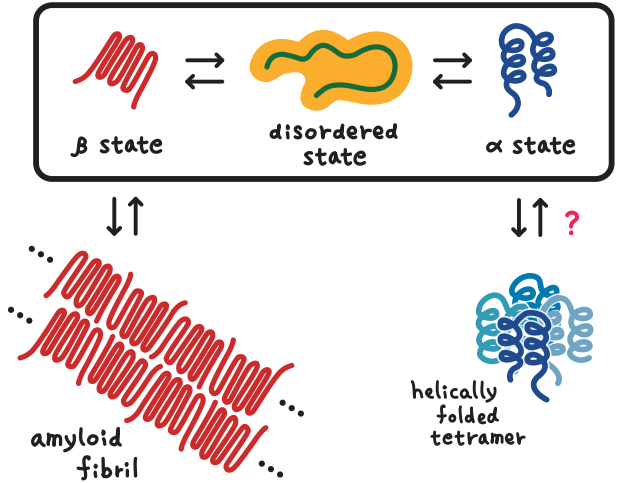}
}
\caption{
Equilibrium of $\alpha$-synuclein. 
Monomeric $\alpha$-synuclein ($\alpha$-syn) is in equilibrium between three types of states: $\beta$-sheet-rich states ($\beta$ states), disordered states and $\alpha$-helix-rich states ($\alpha$ states).
$\alpha$-syn aggregates into a $\beta$-sheet-rich amyloid fibril above a critical concentration. 
The existence of helically folded tetramers is still controversial. 
\label{fig:Equilibrium}
}
\end{figure}

In 2011, Bartels {\it et~al.} reported the existence of helically folded tetramers of $\alpha$-syn~\cite{BartelsSelkoe2011}. 
Although helically folded monomers of $\alpha$-syn had already been identified as a lipid-bound form at that time~\cite{DavidsonGeorge1998,JensenGoedert1998,EliezerBrowne2001,JaoLangen2004,JaoLangen2008,UlmerNussbaum2005,RaoUlmer2010}, this was the first report of a tetrameric state of helically folded monomers. 
These tetramers resist amyloid fibril formation, and they suggested that stabilization of helically folded tetramers is one of the possible strategies for designing an amyloid inhibitor. 
There are several other reports which support the result of Bartels {\it et~al.}~\cite{WangHoang2011,DettmerSelkoe2013,NewmanDettmer2013,WangRoy2014,GouldIschiropoulos2014,DettmerSelkoe2015jun,DettmerSelkoe2015aug}. 
The existence of helically folded tetramers is, however, still controversial; Fauvet {\it et~al.}~\cite{FauvetLashuel2012may} and Burr\'e {\it et~al.}~\cite{BurreSudhof2013} concluded that $\alpha$-syn exists predominantly as a disordered monomer. 
Fauvet {\it et~al.} suggested that stabilization of a disordered monomer of $\alpha$-syn can be an alternative strategy for inhibiting fibril formation. 
We summarize the equilibrium state of $\alpha$-syn including the hypothetical tetrameric states in Fig.~\ref{fig:Equilibrium}. 

Minton discussed the effect of macromolecular crowding on protein folding based on statistical-thermodynamic models~\cite{Minton2000jan,Minton2005feb}. 
After Minton's pioneering studies, Cheung {\it et~al.}~\cite{CheungThirumalai2005} and Kudlay {\it et~al.}~\cite{KudlayThirumalai2009} discussed the same effect based on off-lattice coarse-grained polypeptide models with spherical crowding agents. 
They also analyzed the effect of macromolecular crowding on protein aggregation~\cite{OBrienThirumalai2011}. 
Although they simulated up to four multi-polypeptide systems with crowding agents, this system is not large enough to analyze fibril formation, and thus a coarser-grained model is needed to analyze the effect of macromolecular crowding on the fibril formation. 

Maiti {\it et~al.}~\cite{MaitiSastry2010} proposed a lattice gas model for protein solutions in which three different internal states of a protein are taken into account, and they analyzed competition between protein folding and aggregation. 
Bieler {\it et~al.}~\cite{BielerVacha2012} and \v{S}ari\'{c} {\it et~al.}~\cite{SaricFrenkel2014} constructed off-lattice coarse-grained models for amyloid fibril formation by introducing two internal states, one of which is a fibril-forming $\beta$-sheet state. 
Zhang and Muthukumar~\cite{ZhangMuthukumar2009}, and Irb\"{a}ck {\it et~al.}~\cite{IrbackWallin2013,IrbackWessen2015}, proposed simpler lattice gas models for fibril formation consisting only of a fibril-forming $\beta$-sheet state by considering degrees of freedom of molecular orientations. 

In this paper, in order to investigate the coupled effect of intrinsic disorder and macromolecular crowding on fibril formation, we construct a highly simplified model of $\alpha$-syn with crowding agents based on a lattice gas model. 
Disordered states and fibril-forming $\beta$-sheet-rich states ($\beta$ states) are assumed as internal states of $\alpha$-syn. 
We also investigate the thermal stability of helically folded tetramers by adding tetramer-forming $\alpha$-helix-rich states ($\alpha$ states) to the internal states of $\alpha$-syn. 
Although some reports have discussed the enthalpic effect of crowding agents~\cite{McGuffeeElcock2010,KimMittal2013,HaradaFeig2013}, it is known that thermodynamic properties of biomolecules are significantly altered by considering only the excluded volume effect of crowding agents~\cite{ZhouMinton2008}. 
We only consider this effect in our model as a first step in understanding the fibril formation of $\alpha$-syn. 

\section{Models} \label{sec:model}
\subsection{Construction of the lattice gas model of $\alpha$-synuclein}
\begin{figure}[h]
\centerline{
\includegraphics[width=0.34\textwidth]{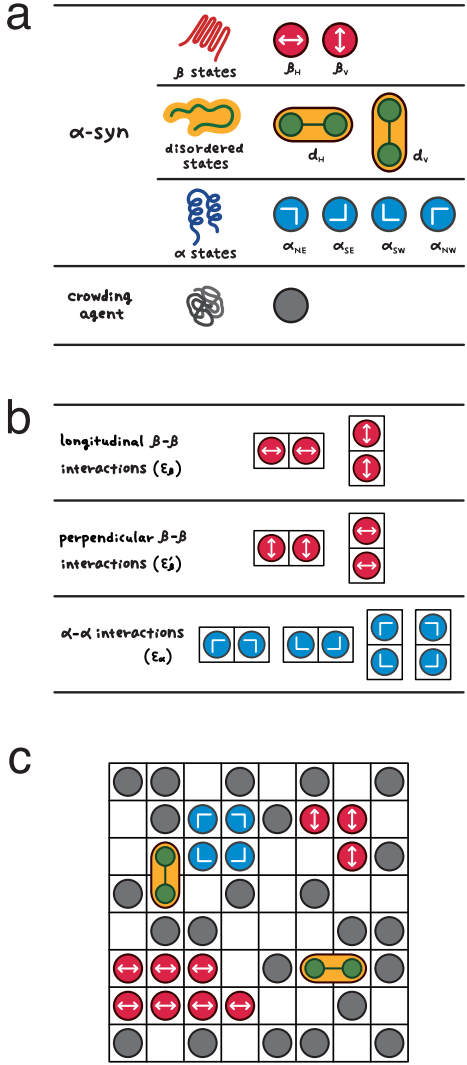}
}
\caption{
Lattice gas model of $\alpha$-synuclein. 
(a) Definitions of two molecules. 
$\alpha$-synuclein ($\alpha$-syn) has three types of internal states: two $\beta$ states ($\beta_H$ and $\beta_V$), two disordered states ($d_H$ and $d_V$) and four $\alpha$ states ($\alpha_{NE}$, $\alpha_{SE}$, $\alpha_{SW}$ and $\alpha_{NW}$). 
$H$ and $V$ denote horizontal and vertical, respectively. 
$N$, $S$, $E$ and $W$ denote north, south, east and west, respectively. 
The crowding agent has only one state. 
$d_H$ and $d_V$ occupy two sites, while the other internal states of $\alpha$-syn and the crowding agent occupy a single site. 
(b) There are three types of interactions between two $\alpha$-syns at nearest-neighboring sites: two longitudinal $\beta$--$\beta$ interactions, two perpendicular $\beta$--$\beta$ interactions and four $\alpha$--$\alpha$ interactions. 
The interaction constants of these three interactions are given by $\varepsilon_\beta$, $\varepsilon^{\prime}_\beta$ and $\varepsilon_\alpha$, respectively. 
(c) Snapshot of a small system ($L=8$, $N=16$) of the lattice gas model. 
\label{fig:aSynStates}
}
\end{figure}

We constructed a model of $\alpha$-syn based on a two-dimensional lattice gas model. 
$\alpha$-syns and crowding agents are modeled as particles on a lattice. 
The definitions of these molecules are shown in Fig.~\ref{fig:aSynStates}(a). 
A single $\alpha$-syn particle has three types of internal states, which have different structures and intermolecular interactions: $\beta$ states, disordered states and $\alpha$ states. 
We further introduce the variations of each type of state which correspond to orientations of $\alpha$-syn: two $\beta$ states ($\beta_H$ and $\beta_V$), two disordered states ($d_H$ and $d_V$) and four $\alpha$ states ($\alpha_{NE}$, $\alpha_{SE}$, $\alpha_{SW}$ and $\alpha_{NW}$).
The meanings of the subscripts are as follows: $H$ and $V$ denote horizontal and vertical, respectively, and $N$, $S$, $E$ and $W$ denote north, south, east and west, respectively. 

We assumed that the disordered states are larger in size than the other states based on the fact that disordered $\alpha$-syn has a larger gyration radius than folded proteins with the same number of residues~\cite{WeinrebLansbury1996,KimYang2002,UverskyFink2001apr}. 
The disordered states occupy two sites, and all the other states and the crowding agents occupy a single site (Fig.~\ref{fig:aSynStates}(a)). 
We also assumed that the disordered states have internal entropy $s$ based on the fact that $\alpha$-syns exhibit a significant entropy loss when they bind to a lipid bilayer surface and fold into an $\alpha$-helical structure~\cite{NuscherBeyer2004,BartelsBeyer2010}. 
We simply introduced the difference of entropy between disordered states and folded states through $s$. 
The idea of the internal entropy was originally introduced by Maiti {\it et~al.}~\cite{MaitiSastry2010} in their three-state lattice gas model. 
In addition to internal entropy, we consider the difference in the volume of each internal state, which is an essential assumption to consider the effect of macromolecular crowding in equilibrium. 

We defined two types of interactions between two $\beta_H$ and two $\beta_V$ based on the model of amyloid fibril formation proposed by Zhang and Muthukumar~\cite{ZhangMuthukumar2009}. 
The definitions of these interactions are shown in Fig.~\ref{fig:aSynStates}(b). 
Two horizontally-aligned $\beta_H$ and two vertically-aligned $\beta_V$ exhibit longitudinal $\beta$--$\beta$ interactions. 
A linear chain of $\alpha$-syns aligned by the longitudinal $\beta$--$\beta$ interactions defines longitudinal direction of a fiber. 
Two vertically-aligned $\beta_H$ and two horizontally-aligned $\beta_V$ exhibit perpendicular $\beta$--$\beta$ interactions. 
The interaction constants of the longitudinal and perpendicular $\beta$--$\beta$ interactions are given by $\varepsilon_\beta$ and $\varepsilon^{\prime}_\beta$ $(\varepsilon_\beta<0, \varepsilon^{\prime}_\beta \le 0, |\varepsilon^{\prime}_\beta|\le |\varepsilon_\beta|)$, respectively.

We defined four $\alpha$--$\alpha$ interactions between four pairs of $\alpha$ states, as shown in Fig.~\ref{fig:aSynStates}(b). 
Dimers, trimers and tetramers of the $\alpha$ states are formed due to $\alpha$--$\alpha$ interactions. 
There are no oligomers of the $\alpha$ states larger than tetramers connected by the $\alpha$--$\alpha$ interactions. 
The interaction constant of the $\alpha$--$\alpha$ interactions is $\varepsilon_\alpha$ ($\varepsilon_\alpha<0$). 

The crowding agents do not have intermolecular interaction except excluded volume interactions. 

By including or excluding the internal states and the interactions of $\alpha$-syn, we defined the following four types of $\alpha$-syn models: $BD$, $BDA$, $B^o \! D$ and $B^o \! DA$.
The latter two are controls for the former two. 
The models with a ``$B$" in their names include the $\beta$ states with both longitudinal and perpendicular $\beta$--$\beta$ interactions. 
Those with a ``$B^o$" include the $\beta$ states which exhibit only the longitudinal $\beta$--$\beta$ interactions. 
Those with a ``$D$" include the disordered states, and those with an ``$A$" include the $\alpha$ states. 

We considered a canonical ensemble of each of the four models presented above, combined with a grand canonical ensemble of crowding agents. 
We defined a system by an $L \times L$ square lattice with $N$ $\alpha$-syns surrounded by walls in contact with a crowding agent reservoir at temperature $T$ and chemical potential $\mu$. 
$\mu$ controls the number of crowding agents. 
The $\mu=-\infty$ system corresponds to the system without crowding agents. 

We show a snapshot of a small system ($L=8$, $N=16$) of the $BDA$ model in Fig.~\ref{fig:aSynStates}(c), which includes all the internal states and all the interactions, as an example. 
A short fiber and a helically folded tetramer are seen in the snapshot. 
The numbers of the internal states of $\alpha$-syn $n_\beta$, $n_d$ and $n_\alpha$ are counted as $n_\beta=10$, $n_d=2$ and $n_\alpha=4$. 
The number of crowding agents is $m=23$. 
The total energy $E$ is calculated as $E = 6\varepsilon_\beta + 4\varepsilon^{\prime}_\beta + 4\varepsilon_\alpha$.

The grand partition function of the $N$ $\alpha$-syn system is given by
\begin{equation}
\Xi = \sum_{E,n_d,m}
W(E,n_d,m)\,
\mathrm{e}^{-\frac{E}{T}} \mathrm{e}^{s n_d} \mathrm{e}^{\frac{\mu}{T}m},
\label{eq:grand_partition_function}
\end{equation}
where $W(O_1,O_2,O_3,\ldots)$ is the number of states of the system with physical quantities $O_i$ ($i=1,2,3,\ldots$). 
We took the Boltzmann constant $k_B=1$.
The thermal average of a physical quantity $O$ is calculated as
\begin{equation}
\langle O \rangle 
= \frac{1}{\Xi} \sum_{O,E,n_d,m} O\, W(O,E,n_d,m)\,
\mathrm{e}^{-\frac{E}{T}}
\mathrm{e}^{s n_d}
\mathrm{e}^{\frac{\mu}{T}m}.
\label{eq:thermal_average_Xi}
\end{equation}

We do not discuss the dynamics of the system, but discuss only the equilibrium states of the system throughout this study. 
We use the following values throughout this paper: $L=128$, $T=1$, $\varepsilon_\beta=-8$, $\varepsilon^{\prime}_\beta=-0.5$, $\varepsilon_\alpha=\varepsilon_\beta/1.2 \simeq -6.67$ and $s=5$. 
The value of $s$ is selected so that the disordered states are predominant as monomeric states at small $\mu$, which is required by the fact that $\alpha$-syns are IDPs~\cite{WeinrebLansbury1996,EliezerBrowne2001,UverskyDunker2008}. 
$\varepsilon_\beta$ and $\varepsilon_\alpha$ were roughly determined so that both $\beta$ and $\alpha$ states are observed evenly in a preliminary simulation for a small system ($B^o \! DA$ model, $L=8$, $N=8$, $T=1$, $\mu=0$, $\sigma=5$).
$\varepsilon^{\prime}_\beta$ was determined so that a protofibril grows unidirectionally. 
We use $N$ and $\mu$ as parameters to investigate their effect on the fibril formation. 
The results for $L=64$ are given in Appendix C for comparison purpose. 
We obtained results similar to those for $L=128$. 

\subsection{Mathematical derivation of effective internal entropy $\sigma$: macromolecular crowding reduces the internal entropy of the disordered states}
In the $N$ $\alpha$-syn system, there are $L^2-N-n_d$ empty sites, and we can count configurations of $m$ single-site crowding agents by the combination
$\begin{pmatrix}
L^2-N-n_d\\
m
\end{pmatrix}$, where $0\le m \le L^2-N-n_d$. 
Using the identity
\begin{equation}
\sum^{L^2-N-n_d}_{m=0} \begin{pmatrix}
L^2-N-n_d\\
m
\end{pmatrix}
\mathrm{e}^{\frac{\mu}{T}m}
= \left( 1+ \mathrm{e}^\frac{\mu}{T} \right)^{L^2-N-n_d}, \notag
\end{equation}
we can rewrite the grand partition function in Eq.~\ref{eq:grand_partition_function} as
\begin{align}
\Xi 
&= \left(1+\mathrm{e}^{\frac{\mu}{T}} \right)^{L^2-N}\, \sum_{E,n_d} w(E,n_d)\, \mathrm{e}^{-\frac{E}{T}} \mathrm{e}^{\sigma n_d} \notag \\
&= \left(1+\mathrm{e}^{\frac{\mu}{T}} \right)^{L^2-N}\, Z, \notag
\end{align}
where $w(O_1,O_2,O_3,\ldots)$ is the number of states of the system with $m=0$ and physical quantities $O_i$ ($i=1,2,3,\ldots$); the canonical partition function $Z$ and the effective internal entropy $\sigma$ are given by
\begin{eqnarray}
Z &=& \sum_{E,n_d} w(E,n_d)\, \mathrm{e}^{-\frac{E}{T}} \mathrm{e}^{\sigma n_d},
\label{eq:partition_function}\\
\sigma &=& s - \log \left( 1+\mathrm{e}^{\frac{\mu}{T}} \right).
\label{eq:sigma}
\end{eqnarray}
If $O$ is not a function of $m$, Eq.~\ref{eq:thermal_average_Xi} is rewritten further using $Z$ as
\begin{equation}
\langle O \rangle
= \frac{1}{Z} \sum_{O,E,n_d} O\, w(O,E,n_d)\, \mathrm{e}^{-\frac{E}{T}} \mathrm{e}^{\sigma n_d}.
\label{eq:thermal_average_Z}
\end{equation}

These equations mean that the original grand canonical ensemble with a crowding agent reservoir is equivalent to a canonical ensemble of an $N$ $\alpha$-syn system without crowding agents. 
In this expression, the effect of macromolecular crowding is absorbed in the effective internal entropy $\sigma$ of the disordered states as the term $-\log\left(1+\mathrm{e}^\frac{\mu}{T}\right)$. 
In the case of $\mu>-\infty$, $\sigma$ is smaller than $s$ by $\log\left(1+\mathrm{e}^\frac{\mu}{T}\right)$, and thus we can say that the crowding effect reduces the internal entropy of the disordered states. 
As long as we do not discuss the dynamics but discuss only the equilibrium states, the newly derived canonical ensemble with effective internal entropy $\sigma$ gives exactly the same results as the grand canonical ensemble with the crowding agents. 
Thus, we successfully expressed the relation between intrinsic disorder and macromolecular crowding in a mathematical form thanks to the simplicity of the model. 

\subsection{Generalization of effective internal entropy $\sigma$}
In the case of $O=m$, Eq.~\ref{eq:thermal_average_Xi} is written as
\begin{align}
\langle m \rangle 
&= T \frac{\partial}{\partial \mu} \log \Xi \notag\\
&= T \frac{\partial}{\partial \mu} \log \left( 1+ \mathrm{e}^\frac{\mu}{T} \right)^{L^2-N} + T \frac{\partial}{\partial \mu} \log Z \notag\\
&= \frac{1}{\mathrm{e}^{- \frac{\mu}{T}}+1}
\Bigl\{ L^2-N-\langle n_d \rangle \Bigr\}, 
\label{eq:thermal_average_of_m}
\end{align}
using the Fermi distribution function $\frac{1}{\mathrm{e}^{- \mu/T}+1}$. 

If we define the volume fraction of $\alpha$-syns $\phi_\mathrm{syn}$ as $\frac{N+ \langle n_d \rangle}{L^2}$ and that of the crowding agents $\phi_m$ as $\frac{\langle m \rangle}{L^2}$, then we can rewrite Eq.~\ref{eq:sigma}
as
\begin{equation}
\sigma 
= s - \log \left(
\frac{1-\phi_\mathrm{syn}}{ 1  - \phi_\mathrm{syn} - \phi_m}
\right). 
\label{eq:before_expansion}
\end{equation}
This form is more suitable to relate to experiments. 
We used values of $N$ ranging from $38$ to $1024$, and thus $\phi_\mathrm{syn}$ ranges from $(38+0)/128^2 \simeq 2.3\times 10^{-3}$ to $(1024+1024)/128^2 \simeq 0.13$. 
Although it is difficult to compare the volume fraction of this two-dimensional model with that of experiments, the values shown above are not so unrealistic. 

\begin{figure*}[t]
\centerline{
\SetLabels
(.025*.9)    \textbf{\huge a}\\
(.53*.9)    \textbf{\huge b}\\
\endSetLabels
\strut\AffixLabels{
\includegraphics[width=0.8\textwidth]{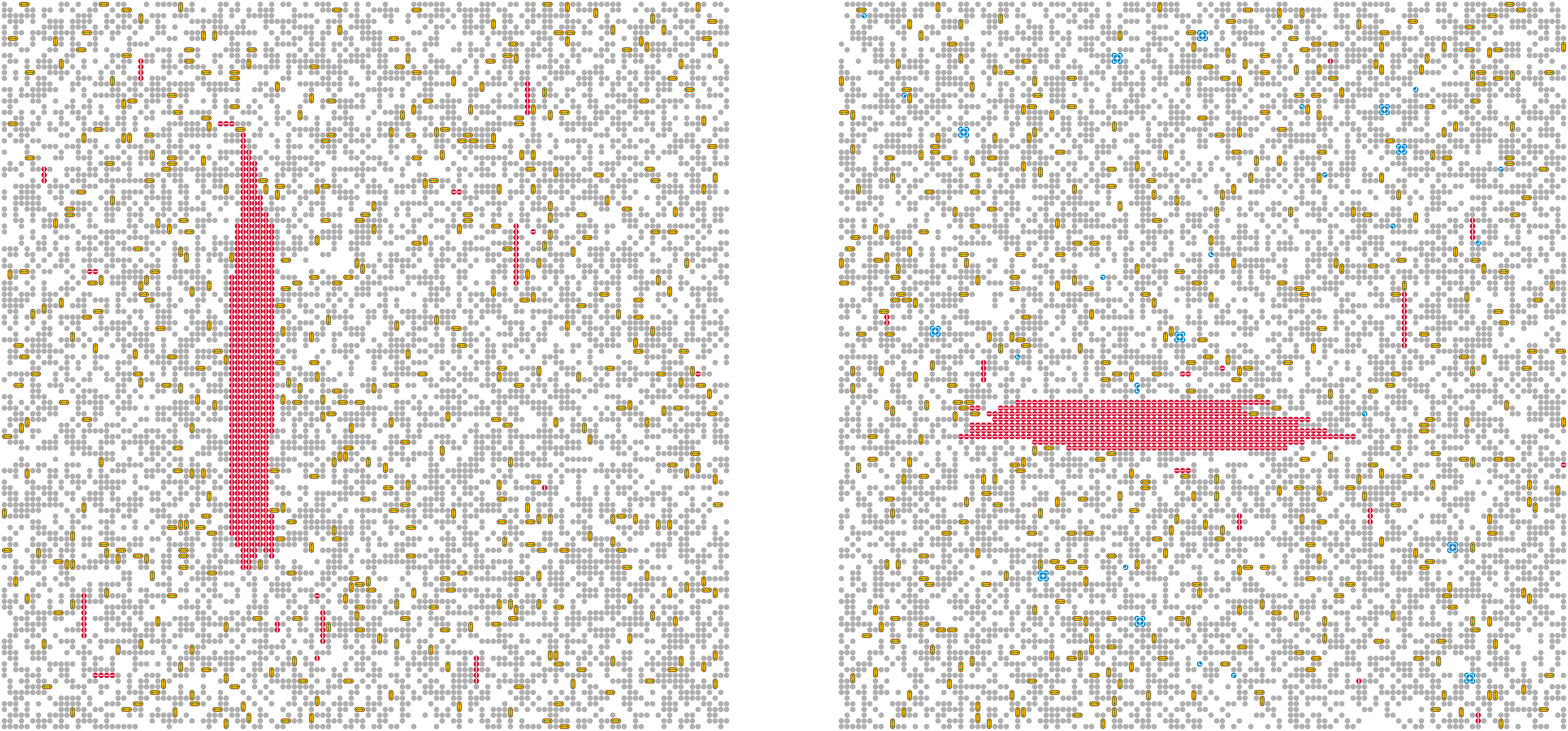}
}
}
\caption{
Protofibril formation of the $\alpha$-synuclein lattice gas model. 
(a) Snapshot for the $BD$ model with $N=1024$ and $\mu=0$.
(b) Snapshot for the $BDA$ model with $N=1024$ and $\mu=0$.
Crowding agents in these snapshots were added after the simulation without explicit crowding agents based on the method shown in Appendix B. 
\label{fig:Snapshots_BD_BDA}
}
\end{figure*}

In this model, we defined the volume of the $\beta$ and $\alpha$ states as a single site, and that of the disordered states as double sites. 
Thus, the volume difference between the disordered states and the other states is $1$. 
For the case in which the volume difference is $\Delta v$, we can also derive the general form of Eqs.~\ref{eq:sigma}
and \ref{eq:before_expansion} as
\begin{align}
\sigma &= s - \Delta v \, \log \left( 1+\mathrm{e}^{\frac{\mu}{T}} \right),
	\label{eq:sigma2} \\
\sigma &= s - \Delta v \, \log \left(
	\frac{1-\phi_\mathrm{syn}}{ 1  - \phi_\mathrm{syn} - \phi_m}
	\right).
	\label{eq:before_expansion2}
\end{align}
This equation is applicable as long as the volume of the folded states is greater than or equal to that of the crowding agents. 
Eq.~\ref{eq:before_expansion2} will be a first approximation to discuss the relation between macromolecular crowding and the stability of disordered states also for larger crowding agents. 
Eq.~\ref{eq:before_expansion2} provides a simple theoretical expression for the interplay of intrinsic disorder and macromolecular crowding, which was suggested by White {\it et~al.}~\cite{WhiteDobson2010}. 

Note that the same analysis is valid also for three-dimensional lattice models as long as we consider single-site crowding agents. 

\section{Results}
\subsection{Protofibril formation above the critical concentration of $\alpha$-synuclein}
We calculate thermal averages through a Monte Carlo simulation using the Metropolis method based on Eq.~\ref{eq:partition_function}. 
We do not need to consider crowding agents explicitly in the simulation because we introduce $\sigma$ calculated from Eq.~\ref{eq:sigma} with a given $\mu$. 
Details of the method are given in Appendix A.

Before introducing the $\alpha$ states, we focus on the equilibrium between the $\beta$ states and the disordered states using the $BD$ model. 
Fig.~\ref{fig:Snapshots_BD_BDA}(a) shows snapshots of the $BD$ model with $N=1024$ and $\mu=0$.  
There is a two-dimensional $\beta$-state cluster, which involves both longitudinal and perpendicular $\beta$--$\beta$ interactions. 
We refer to the $\beta$-state cluster as a protofibril. 

First, we analyze the effect of $\alpha$-syn concentration on protofibril formation.
We refer to a linear chain of $\beta$ states connected by longitudinal $\beta$--$\beta$ interactions as a $\beta$ chain, and we define $\ell$ as the length of a $\beta$ chain. 
If a $\beta$ state is not connected to other $\alpha$-syns by longitudinal $\beta$--$\beta$ interactions, we count $\ell$ as 1. 
We define $h(\ell)$ as the number of $\beta$ chains of length $\ell$. 
Fig.~\ref{fig:Length_plot}(a) shows $N$ dependences of the distributions of $\langle \ell \cdot h(\ell)\rangle$ against $\ell$ for the $BD$ and $B^o \! D$ models with $\mu=0$. 
The pairs of the curves of two models coincide with each other for the $N=512$ and $609$ systems. 
They separate in the $N=724$ and $861$ systems, and the curves for the $BD$ model have two peaks; we found from snapshots that $\beta$ chains with $\ell$ around the right peaks form a protofibril.
We observed formation and dissolution of a protofibril for $N=724$. 
Thus, an increase in the $\alpha$-syn concentration induces the protofibril formation and there is a critical value of $N$ around $724$. 
We write the critical value as $N_c$. 
For $N>N_c$, even if there are multiple protofibrils in the early stage of the simulation, they merge into a single thermally stable protofibril. 
The existence of $N_c$ qualitatively agrees with the experiments of $\alpha$-syn fibril formation~\cite{HashimotoMasliah1998,RaaijSubramaniam2008,ShtilermanLansbury2002,UverskyFink2002}. 

\subsection{Macromolecular crowding induces protofibril formation}
\begin{figure}[h]
\centerline{
\includegraphics[width=0.35\textwidth]{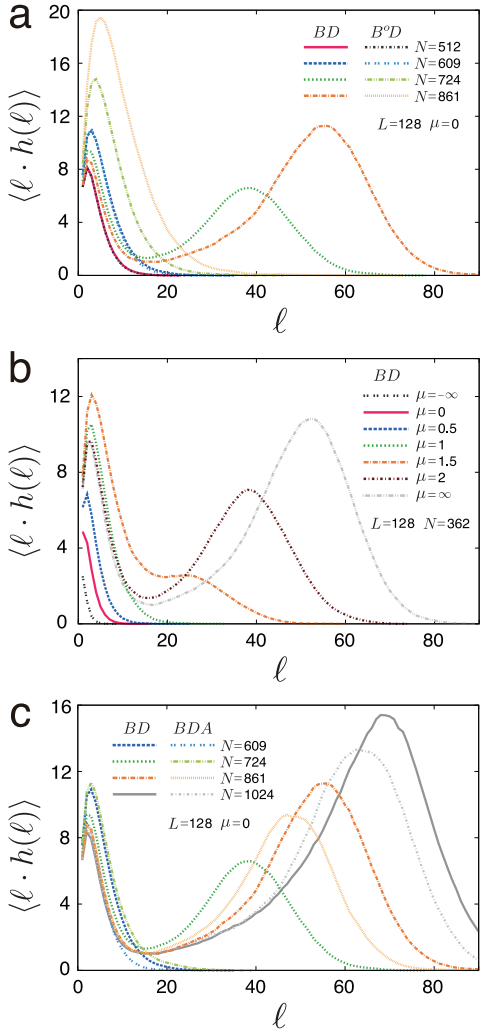}
}
\caption{
Equilibrium number of $\beta$ states which are components of $\beta$ chains of length $\ell$. 
We define a $\beta$ chain as a linear chain of $\beta$ states connected by longitudinal $\beta$--$\beta$ interactions. 
(a) $N$ dependences of the $\langle \ell \cdot h(\ell)\rangle$ distributions against $\ell$ for the $BD$ and $B^o \! D$ models. 
(b) $\mu$ dependences of the $\langle \ell \cdot h(\ell)\rangle$ distributions against $\ell$ for the $BD$ model. 
(c) $N$ dependences of the $\langle \ell \cdot h(\ell)\rangle$ distributions against $\ell$ for the $BD$ and $BDA$ models. 
\label{fig:Length_plot}
}
\end{figure}

Next, we analyze the effect of macromolecular crowding on protofibril formation. 
Fig.~\ref{fig:Length_plot}(b) shows 
the distributions of $\langle \ell \cdot h(\ell) \rangle$ against $\ell$ for the $BD$ model with $N=362$. 
As $\mu$ increases, the right peak begins to grow at $\mu=1.5$, and gets larger at $\mu=2$. 
We observed formation and dissolution of a protofibril in these systems. 
Thus, macromolecular crowding induces protofibril formation, and the critical value of the chemical potential $\mu_c$ is around $1.5$. 
This qualitatively agrees with the experiments for $\alpha$-syn fibril formation with crowding agents~\cite{ShtilermanLansbury2002,UverskyFink2002}. 

\begin{figure}[h]
\centerline{
\includegraphics[width=0.35\textwidth]{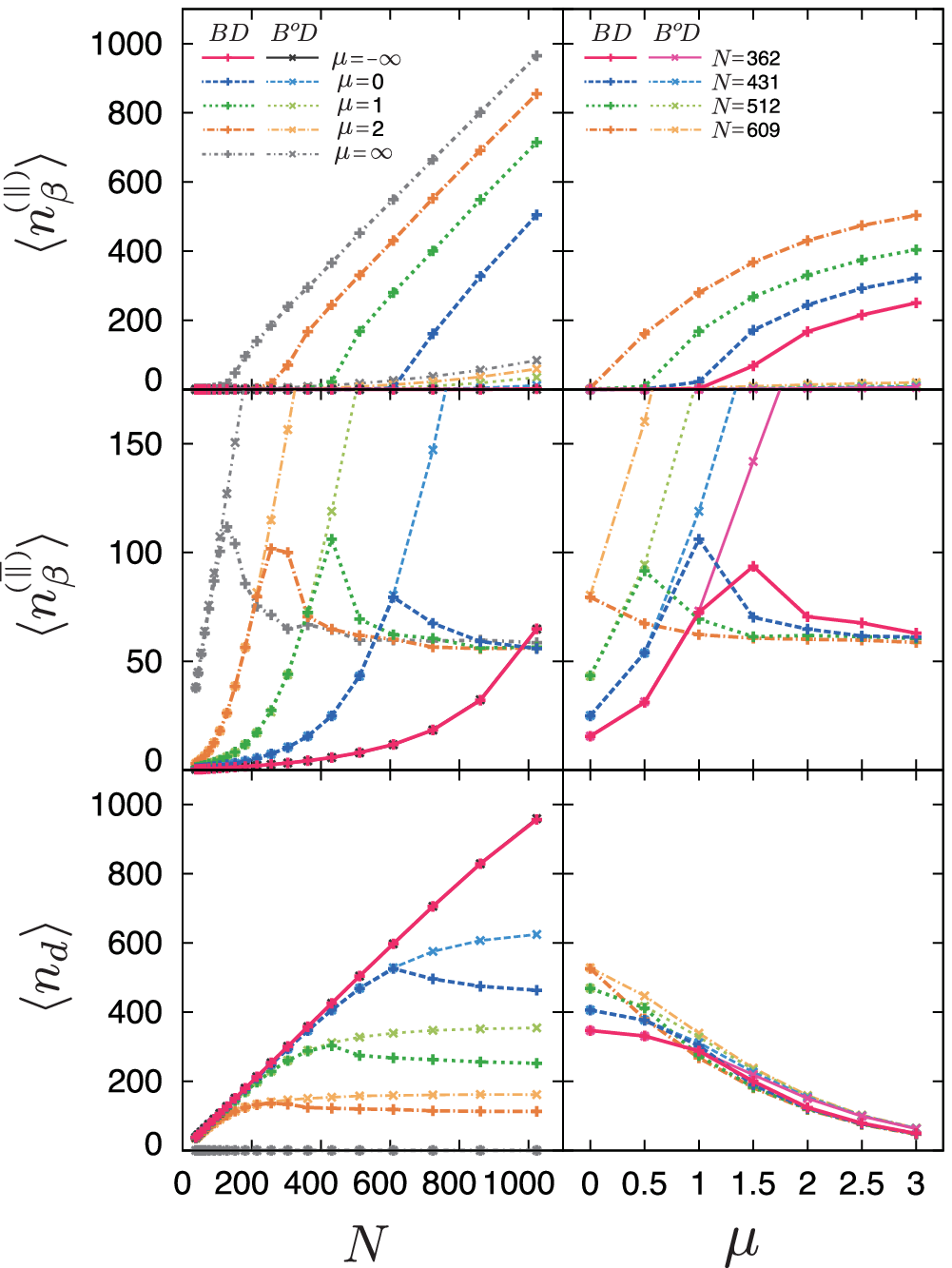}
}
\caption{
$N$ and $\mu$ dependences of the numbers of $\alpha$-synuclein states for the $BD$ and $B^o \! D$ models. 
\label{fig:BD_ave_multiplot}
}
\end{figure}

We define $n^{(\parallel)}_\beta$ as the number of the $\beta$ states connected to other $\beta$ states by perpendicular $\beta$--$\beta$ interactions and $n^{(\overline{\parallel})}_\beta$ as the number of other $\beta$ states.
The left column of Fig.~\ref{fig:BD_ave_multiplot} shows $\langle n^{(\parallel)}_\beta \rangle$, $\langle n^{(\overline{\parallel})}_\beta \rangle$ and $\langle n_d \rangle$ against $N$ for the $BD$ and $B^o \! D$ models with fixed $\mu$. 
$\mu$ was set as $-\infty$, $0$, $1$, $2$ and $\infty$. 
We counted the number of horizontally aligned $\beta_V$ and vertically aligned $\beta_H$ as $n^{(\parallel)}_\beta$ also for the $B^o \! D$ and $B^o \! DA$ models in which perpendicular $\beta$--$\beta$ interactions are absent. 
In all these three graphs, the curves for the $BD$ and $B^o \! D$ models with the same $\mu$ separate at $N_c$ except for $\mu=\infty$. 
The slopes of $\langle n^{(\parallel)}_\beta \rangle$ for the $BD$ model change from $\sim 0$ to $\sim 1$ around $N_c$. 
$\langle n^{(\overline{\parallel})}_\beta \rangle$ for the $BD$ model has a maximum around $N_c$, and decreases to the constant value which does not depend on $\mu$ for large $N$. 
$\langle n_d \rangle$ for the $BD$ model also has a maximum around $N_c$, and decreases to the constant value which depends on $\mu$. 
We found that $N_c$ decreases as $\mu$ increases by comparing the separation points. 
This means that macromolecular crowding reduces the critical concentration of $\alpha$-syn, and this also means that macromolecular crowding induces protofibril formation. 

The right column of Fig.~\ref{fig:BD_ave_multiplot} shows $\langle n^{(\parallel)}_\beta \rangle$, $\langle n^{(\overline{\parallel})}_\beta \rangle$ and $\langle n_d \rangle$ against $\mu$ for the $BD$ and $B^o \! D$ models with fixed $N$. 
$N$ was set as $362$, $431$, $512$ and $609$. 
In all three graphs of Fig.~\ref{fig:BD_ave_multiplot}, the curves for the $BD$ and $B^o \! D$ models with the same $N$ separate at $\mu_c$. 
$\langle n^{(\overline{\parallel})}_\beta \rangle$ for the $BD$ model has a maximum around $\mu_c$, and decreases to the constant value which does not depend on $N$ for large $\mu$. 
$\langle n_d \rangle$ for both models monotonically decreases along with $\mu$. 
On the other hand, $\langle n_\beta \rangle$ for both models monotonically increases along with $\mu$ because $\langle n_\beta \rangle = N-\langle n_d \rangle$. 
Thus, macromolecular crowding suppresses the disordered states and promotes the $\beta$ states, which reflects the reduction of the effective internal entropy $\sigma$. 
We found that $\mu_c$ decreases as $N$ increases by comparing the separation points for different values of $N$. 
This means that $\alpha$-syn concentration reduces the critical chemical potential of crowding agents, and this again means that macromolecular crowding induces protofibril formation. 

\subsection{$\alpha$-state tetramers suppress protofibril formation}
We introduce $\alpha$ states and investigate the effect of helically folded tetramers on the protofibril formation.
Fig.~\ref{fig:Length_plot}(c) shows
the distributions of $\langle \ell \cdot h(\ell) \rangle$ against $N$ for the $BD$ and $BDA$ models with $\mu=0$. 
As we have seen in Fig.~\ref{fig:Length_plot}(a), the curve of the $BD$ model for $N=724$ has two peaks and $N_c$ for the $BD$ model is around $724$. 
The corresponding curve of the $BDA$ model, however, has only a single peak and thus $N_c$ is higher than that of the $BD$ model. 
This means that introducing of $\alpha$ states raises $N_c$; in other words, $\alpha$ states suppress protofibril formation. 

\begin{figure}[h]
\centerline{
\includegraphics[width=0.3\textwidth]{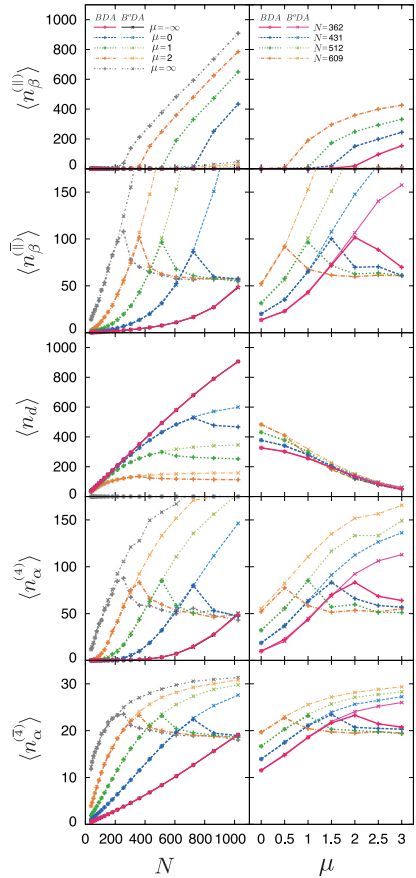}
}
\caption{
$N$ and $\mu$ dependences of the numbers of $\alpha$-synuclein states for the $BDA$ and $B^o \! DA$ models. 
\label{fig:BDA_ave_multiplot}
}
\end{figure}

A protofibril coexists with short $\beta$ chains and tetramers in the snapshot for the $BDA$ model with $N=1024$ (Fig.~\ref{fig:Snapshots_BD_BDA}(b)).
To analyze the oligomeric states of the $\alpha$ states, we define $\langle n^{(4)}_\alpha \rangle$ as the number of $\alpha$ states forming tetramers, and $\langle n^{(\overline{4})}_\alpha \rangle$ as the number of other $\alpha$ states. 
The left and right columns of Fig.~\ref{fig:BDA_ave_multiplot} show $\langle n^{(\parallel)}_\beta \rangle$, $\langle n^{(\overline{\parallel})}_\beta \rangle$, $\langle n_d \rangle$, $\langle n^{(4)}_\alpha \rangle$ and $\langle n^{(\overline{4})}_\alpha \rangle$ against $N$ and $\mu$, respectively. 
The overall shapes of $\langle n^{(\parallel)}_\beta \rangle$, $\langle n^{(\overline{\parallel})}_\beta \rangle$ and $\langle n_d \rangle$ against $N$ in Fig.~\ref{fig:BDA_ave_multiplot} are similar to those of corresponding graphs in Fig.~\ref{fig:BD_ave_multiplot}. 
All of $\langle n^{(4)}_\alpha \rangle$ and $\langle n^{(\overline{4})}_\alpha \rangle$ for the $BDA$ and $B^o \! DA$ models behave similarly to $\langle n^{(\overline{\parallel})}_\beta \rangle$; each curve has a maximum around $N_c$, and decreases to the constant value which does not depend on $\mu$. 
We found that the majority of $\alpha$ states form tetramers around $N_c$ by comparing $\langle n^{(4)}_\alpha \rangle$ and $\langle n^{(\overline{4})}_\alpha \rangle$. 
Thus, there is a major contribution to the suppression of protofibril formation by $\alpha$-state tetramers, and we can say that $\alpha$-state tetramers suppress fibril formation. 
This is consistent with the report of Bartels {\it et~al.}~\cite{BartelsSelkoe2011}

We found that $N_c$ and $\mu_c$ decrease as $\mu$ and $N$ increase, respectively, by comparing the separation points. 
These behaviors are similar to those observed in the $BD$ model. 
We also found that $N_c$ and $\mu_c$ for the $BDA$ model are larger than those for the $BD$ model. 
This means that introduction of $\alpha$ states raises both the critical concentration of $\alpha$-syn and the critical chemical potential of crowding agents. 

\section{Summary and Discussion}
\subsection{Scenario of protofibril formation induced by macromolecular crowding}
\begin{figure*}[t]
\centerline{
\includegraphics[width=0.9\textwidth]{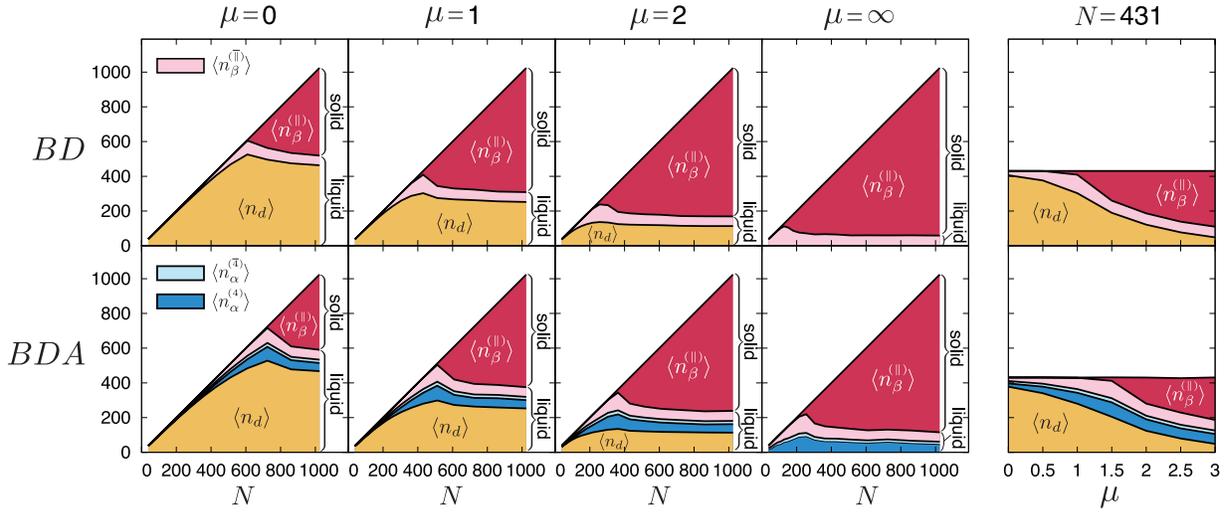}
}
\caption{
Stacked graphs of the equilibrium numbers of $\alpha$-synuclein states for the $BD$ and $BDA$ models. 
$\langle n^{(\parallel)}_\beta \rangle$, $\langle n^{(\overline{\parallel})}_\beta \rangle$, $\langle n^{(\overline{4})}_\alpha \rangle$, $\langle n^{(4)}_\alpha \rangle$ and $\langle n_d \rangle$ are colored red, pink, light blue, blue and yellow, respectively. 
}
\label{fig:stacked_graph}
\end{figure*}

The curves for the $BD$ and $BDA$ models in Figs.~\ref{fig:BD_ave_multiplot} and \ref{fig:BDA_ave_multiplot} are summarized in Fig.~\ref{fig:stacked_graph} as stacked graphs. 
$N$ dependences are shown in the left, and $\mu$ dependences are shown in the right. 
The $\langle n^{(\parallel)}_\beta \rangle$ regions start to increase around $N_c$ in the left graphs and $\mu_c$ in the right graphs, respectively. 
The existence of these critical values qualitatively agrees with the experiments of $\alpha$-syn~\cite{HashimotoMasliah1998,RaaijSubramaniam2008,ShtilermanLansbury2002,UverskyFink2002}. 

The left graphs in Fig.~\ref{fig:stacked_graph} show that $\langle n^{(\parallel)}_\beta \rangle$ monotonically increases above $N_c$, while $\langle n^{(\overline{\parallel})}_\beta \rangle$, $\langle n_d \rangle$, $\langle n^{(4)}_\alpha \rangle$ and $\langle n^{(\overline{4})}_\alpha \rangle$ each decrease to constant values. 
In other words, the protofibril absorbs increased $\alpha$-syns above $N_c$ while the other non-protofibril components converge to a constant concentration. 
From this result, we can draw an analogy of protofibril--non-protofibril equilibrium with a solid--liquid equilibrium; the protofibril can be seen as solid, and the non-protofibril components can be seen as liquid. 
The words ``solid" and ``liquid" are written in the stacked graphs. 

\begin{figure}[h]
\centerline{
\includegraphics[width=0.4\textwidth]{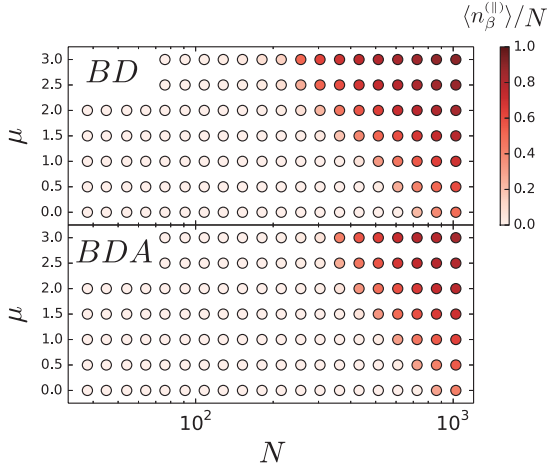}
}
\caption{
Phase diagrams for $BD$ and $BDA$ models. 
Colors indicate the values of $\langle n^{(\parallel)}_\beta \rangle/N$.
\label{fig:phase_diagram}
}
\end{figure}

The phase diagrams shown in Fig.~\ref{fig:phase_diagram} are helpful to see a first-order-like transition between liquid phase and solid-liquid coexistence. 
The upper and lower phase diagrams correspond to the $BD$ and $BDA$ model, respectively. 
The colors of each circle indicate the values of $\langle n_\beta^{(\parallel)} \rangle/N$. 
Each phase diagram has two phases: a liquid phase in the lower-left white region, and solid-liquid coexistence in the upper-right red region. 
$N_c$ and $\mu_c$ form a boundary between the two phases. 
The white region in the phase diagram of the $BDA$ model is larger than that of $BD$ model, and this means that the $BDA$ model has larger $N_c$ and $\mu_c$ than the $BD$ model. 

\begin{figure}[h]
\centerline{
\includegraphics[width=0.45\textwidth]{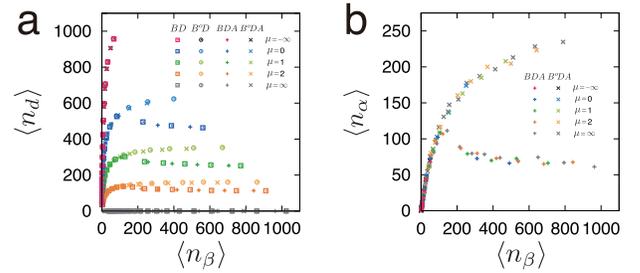}
}
\caption{
$\langle n_\beta \rangle$ dependences of $\alpha$-synuclein states. 
(a) $\langle n_\beta \rangle$ dependences of $\langle n_d \rangle$ of the $BD$, $B^o \! D$, $BDA$ and $B^o \! DA$ models. 
(b) $\langle n_\beta \rangle$ dependences of $\langle n_\alpha \rangle$ for the $BDA$ and $B^o \! DA$ models. 
\label{fig:n_beta_multiplot}
}
\end{figure}

$\langle n_d \rangle$ and $\langle n_\alpha \rangle$ against $\langle n_\beta \rangle$ for the $BDA$ and $B^o \! DA$ models are plotted in Fig.~\ref{fig:n_beta_multiplot}(a) and (b). 
$\langle n_d \rangle$ for the $BD$ and $BDA$ models collapse onto a single curve for each $\mu$, and those for the $B^o \! D$ and $B^o \! DA$ models also do so. 
Thus, $\langle n_d \rangle$ is a function of $\langle n_\beta \rangle$ and $\mu$ irrespective of $\alpha$ states. 
$\langle n_\alpha \rangle$ for all values of $\mu$ collapse onto a single curve for each of the $BDA$ and $B^o \! DA$ models. 
Thus, $\langle n_\alpha \rangle$ is the function of a single variable $\langle n_\beta \rangle$, which is independent of $\mu$. 
We find the single critical value of $\langle n_\beta \rangle$ for the $BD$ and $BDA$ models, and we write it as $\langle n_\beta \rangle_c$. 
Thus, $\langle n_\beta \rangle$ is a good parameter to describe protofibril formation. 

We propose the following scenario for protofibril formation induced by macromolecular crowding based on the stacked graphs for $\mu$ dependence of the internal states presented on the right of Fig.~\ref{fig:stacked_graph}. 
Since the stability of the disordered states is determined by $\sigma$, which is given by Eq.~\ref{eq:sigma}, the $\langle n_d \rangle$ regions decrease along with $\mu$ as a result of decreasing $\sigma$. 
In other words, macromolecular crowding destabilizes the disordered states by reducing the effective internal entropy. 
The $\langle n_\beta \rangle \left(=\langle n_\beta^{(\parallel)} \rangle + \langle n_\beta^{(\overline{\parallel})} \rangle\right)$ region increases as a result of the decrease in $\langle n_d \rangle$ because $\langle n_\beta \rangle = N - \langle n_d \rangle$ in the $BD$ model and $\langle n_\beta \rangle = N - \langle n_d \rangle - \langle n_\alpha \rangle$ in the $BDA$ model. 
When $\langle n_\beta \rangle$ exceeds $\langle n_\beta \rangle_c$, which is the same value for both $BD$ and $BDA$ models, the $\langle n^{(\parallel)}_\beta \rangle$ regions start to increase, and the $\beta$ states form a protofibril. 
At this point, $\mu \simeq \mu_c$. 
Since $\langle n_d \rangle$ is not a function of $\langle n_\alpha \rangle$, as we explained based on Fig.~\ref{fig:n_beta_multiplot}, $\langle n_\beta \rangle$ of the $BDA$ model is smaller than that of the $BD$ model by $\langle n_\alpha \rangle$ for the same $N$ and $\mu$.
Thus, the $BDA$ model has higher $N_c$ and $\mu_c$.
This explains that $\alpha$ states suppress the protofibril formation. 

\subsection{An explanation for the controversy over the observation of helically folded tetramers}
\begin{figure}[h]
\centerline{
\includegraphics[width=0.3\textwidth]{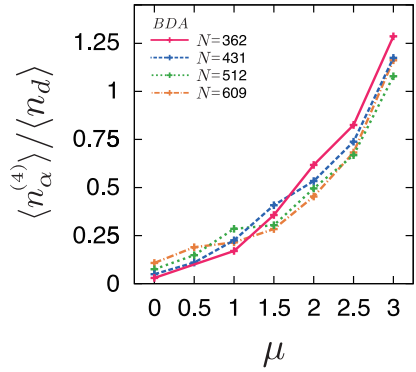}
}
\caption{
$N$ and $\mu$ dependences of the ratio between $\langle n^{(4)}_\alpha \rangle$ and $\langle n_d \rangle$ for the $BDA$ and $B^o \! DA$ models.
\label{fig:tetramer_disorder_ratio}
}
\end{figure}

We consider that the detectability of helically folded tetramers in experiments depends on the abundance of other states, disordered states in particular. 
To discuss their detectability, we plotted the $\mu$ dependences of $\langle n^{(4)}_\alpha \rangle / \langle n_d \rangle$ for the $BDA$ model in Fig.~\ref{fig:tetramer_disorder_ratio}. 
The $\langle n^{(4)}_\alpha \rangle/\langle n_d \rangle$ curves monotonically increase along with $\mu$. 
Thus, macromolecular crowding increases the relative number of helically folded tetramers to that of disordered monomers. 
The macromolecular crowding effect may provide a good explanation for the controversy over the existence of helically folded tetramers. The present result that the density of such tetramers depends on crowdedness suggests that whether or not the tetramers are observed in experiments is determined by the crowdedness of the environment.

\begin{acknowledgments}
This work was supported by a Grant-in-Aid (Grant No. 21113006) from Ministry of Education, Culture, Sports, Science, and Technology (MEXT), and the Global COE Program Core Research and Engineering of Advanced Materials-Interdisciplinary Education Center for Materials Science (Grant No. G10) from Japan Society for the Promotion of Science (JSPS).
\end{acknowledgments}

\appendix

\section*{Appendix A: Metropolis method for the lattice gas model of $\alpha$-synuclein including crowding agents}
\label{subsec:metropolis1}
We used the Metropolis method to produce an equilibrium ensemble.
In the lattice model of $\alpha$-syn, a state of the system is given by conformations of $N$ $\alpha$-syns and $m$ crowding agents on the $L\times L$ lattice. 
Since we do not discuss the dynamics of the system, we can introduce unrealistic moves which fasten the exploration of the configuration space. 
Although crowding agents become obstacles to other molecules and they obviously change the dynamics of the system, we do not have to include them explicitly as long as their entropic effects are properly considered in the calculation. 
We can also use Eq.~\ref{eq:thermal_average_Z}
, which does not include the explicit crowding agents, to calculate thermal averages in place of Eq.~\ref{eq:thermal_average_Xi}
. Using the effective internal entropy $\sigma$, which is written as Eq.~\ref{eq:sigma}
, the
transition probability from state $i$ to state $j$ is given by
\begin{equation}
p(i \to j) =
\min\left[
\frac{\exp\left(-E^j/T+\sigma n^j_d\right)}{\exp\left(-E^i/T+\sigma n^i_d\right)},1
\right].
\label{eq:transition_probability}
\tag{A1}
\end{equation}
We defined the following four Monte Carlo moves for $\alpha$-syns which are performed with the transition probability in Eq.~\ref{eq:transition_probability}. 

\begin{enumerate}
\item[(i)]
Local displacement of a single $\alpha$-syn: 
(a) If a selected $\alpha$-syn is in a $\beta$ or $\alpha$ state, move it in a randomly selected direction (north, south, east or west) by one site. 
This move will be rejected if the destination site is covered with another $\alpha$-syn. 
(b) If a selected $\alpha$-syn is in a disordered state, move it to a randomly selected direction by one site. 
This move will be rejected if one or both of the two destination sites are covered with other $\alpha$-syns. \\
\item[(ii)]
Local displacement of a group of $\alpha$-syns: 
(a) In the case that the selected $\alpha$-syn is a part of a $\beta$-state linear chain connected by longitudinal $\beta$--$\beta$ interactions, move the whole linear chain to a randomly selected direction by one site if the destination sites are empty. 
This move will be rejected if the number of longitudinal $\beta$--$\beta$ interactions increases. 
(b) In the case that the selected $\alpha$-syn is a part of an $\alpha$-state oligomer (dimer, trimer or tetramer) connected by $\alpha$--$\alpha$ interactions, move the oligomer to a randomly selected direction by one site if the destination sites are empty. 
This move will be rejected if the number of $\alpha$--$\alpha$ interactions increases. 
(c) In the case that a selected $\alpha$-syn is not connected with other $\alpha$-syns by longitudinal $\beta$--$\beta$ or $\alpha$--$\alpha$ interaction, perform the move shown in (i), above. 
This move will be rejected if the number of longitudinal $\beta$--$\beta$ or $\alpha$--$\alpha$ interactions increases.\\
\item[(iii)]
Nonlocal displacement of a single $\alpha$-syn: 
(a) If a selected $\alpha$-syn is in a $\beta$ or $\alpha$ state, move it to a randomly selected site of the system. 
This move will be rejected if the destination site is covered with other $\alpha$-syns. 
(b) If a selected $\alpha$-syn is in a disordered state, move it to two randomly selected adjacent sites of the system. 
This move will be rejected if one or both of the two destination sites are covered with other $\alpha$-syns.\\
\item[(iv)]
Internal state switching: 
Switch the internal state of a selected $\alpha$-syn to another randomly selected state. 
This move will be rejected if a disordered state is selected as the next state and if it overlaps with another $\alpha$-syn. 
\end{enumerate}

We randomly select an $\alpha$-syn from the system, and randomly perform one of the moves (i)--(iv) with prescribed probabilities.
In all cases of (i), (ii) and (iii), the move will be rejected if the selected $\alpha$-syns go outside the system. 

\renewcommand{\thefigure}{C1}

\begin{figure*}[t]
\centerline{
\SetLabels
(.025*.9)    \textbf{\huge a}\\
(.53*.9)    \textbf{\huge b}\\
\endSetLabels
\strut\AffixLabels{
\includegraphics[width=0.75\textwidth]{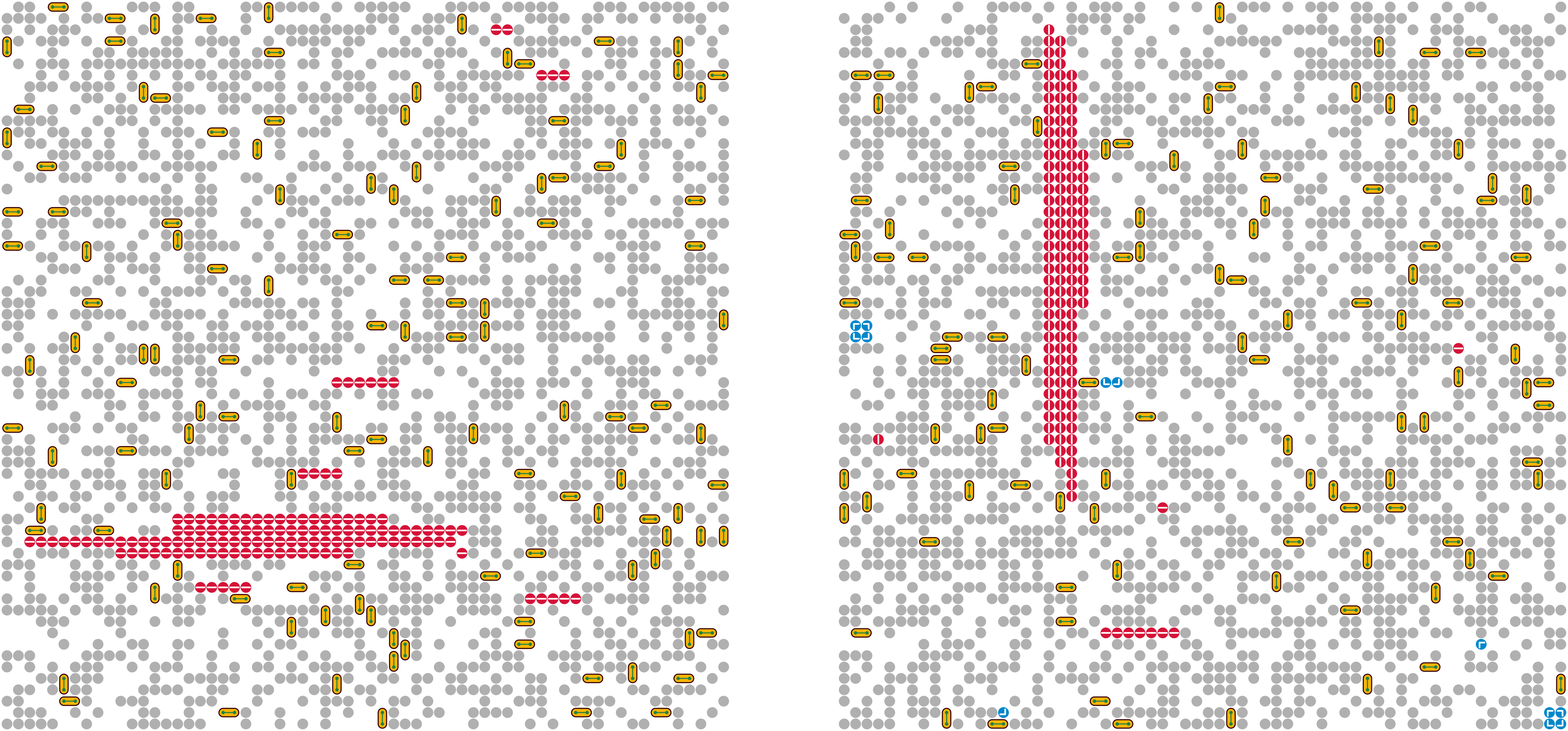}
}
}
\caption{
Snapshots for the systems with $L=64$. 
(a) The $BD$ model and (b) the $BDA$ model with $N=256$ and $\mu=0$.
\label{fig:Snapshots_BD_BDA_L64}
Crowding agents in these snapshots were added after the simulation without explicit crowding agents based on the method shown in Appendix B. 
}
\end{figure*}

\section*{Appendix B: Method of adding crowding agents to the snapshots produced by the simulation without explicit crowding agents}
\label{subsec:metropolis2}
We explain the method of adding crowding agents based on $\mu$ to the snapshots produced by the method shown above. 
We also use the Metropolis method to produce grand canonical ensemble of crowding agents. 
The transition probability from state $i$ to state $j$ is given by
\begin{equation}
p(i \to j) = 
\min\left[
\frac{\exp(\mu \, m^j /T)}{\exp(\mu \, m^i /T)},1
\right].
\tag{B1}
\end{equation}
We randomly select a site from the system and perform the following move: 
\begin{itemize}
\item Insertion or removal of a crowding agent: 
(a) In the case that a selected site is empty, insert a new crowding agent into that site. 
(b) In the case that a selected site is covered by a crowding agent, remove it from the system. 
(c) In the case that a selected site is covered by an $\alpha$-syn, this move will be rejected. 
\end{itemize}
When we created the snapshots shown in Fig.~\ref{fig:Snapshots_BD_BDA}(a) and (b), we first created the snapshots by the simulation without explicit crowding agents, and then we added crowding agents to the snapshots by using the method explained here.

\section*{Appendix C: Brief description of the results for $L=64$}
The results which we have discussed in the main text were from the calculations for the $L=128$ systems. 
Here, we show the results for the $L=64$ systems. 
Fig.~\ref{fig:Snapshots_BD_BDA_L64}(a) and Fig.~\ref{fig:Snapshots_BD_BDA_L64}(b) show snapshots of the $BD$ and $BDA$ models with $L=64$, $N=256$ and $\mu=0$. 
$N/L^2$ of these systems is the same as that of the systems shown in Figs.~\ref{fig:Snapshots_BD_BDA}{\it a} and \ref{fig:Snapshots_BD_BDA}{\it b}. 
In the snapshots of the $L=64$ systems, there is a single protofibril. 
Fig.~\ref{fig:BD_ave_multiplot_L64}, Fig.~\ref{fig:BDA_ave_multiplot_L64} and Fig.~\ref{fig:tetramer_disorder_ratio_L64} show the same physical quantities for the $L=64$ systems as those shown in 
Figs.~\ref{fig:BD_ave_multiplot}, \ref{fig:BDA_ave_multiplot}, and \ref{fig:tetramer_disorder_ratio}, respectively. 
The overall shapes of the curves for the $L=64$ systems are similar to those of the curves for the $L=128$ systems. 
The $L=64$ systems also have $N_c$ and $\mu_c$, and the existence of the $\alpha$-state tetramers increases these values.  
The results of the $L=64$ systems are qualitatively same as those of the $L=128$ systems.

\renewcommand{\thefigure}{C2}
\begin{figure}[h]
\centerline{
\includegraphics[width=0.35\textwidth]{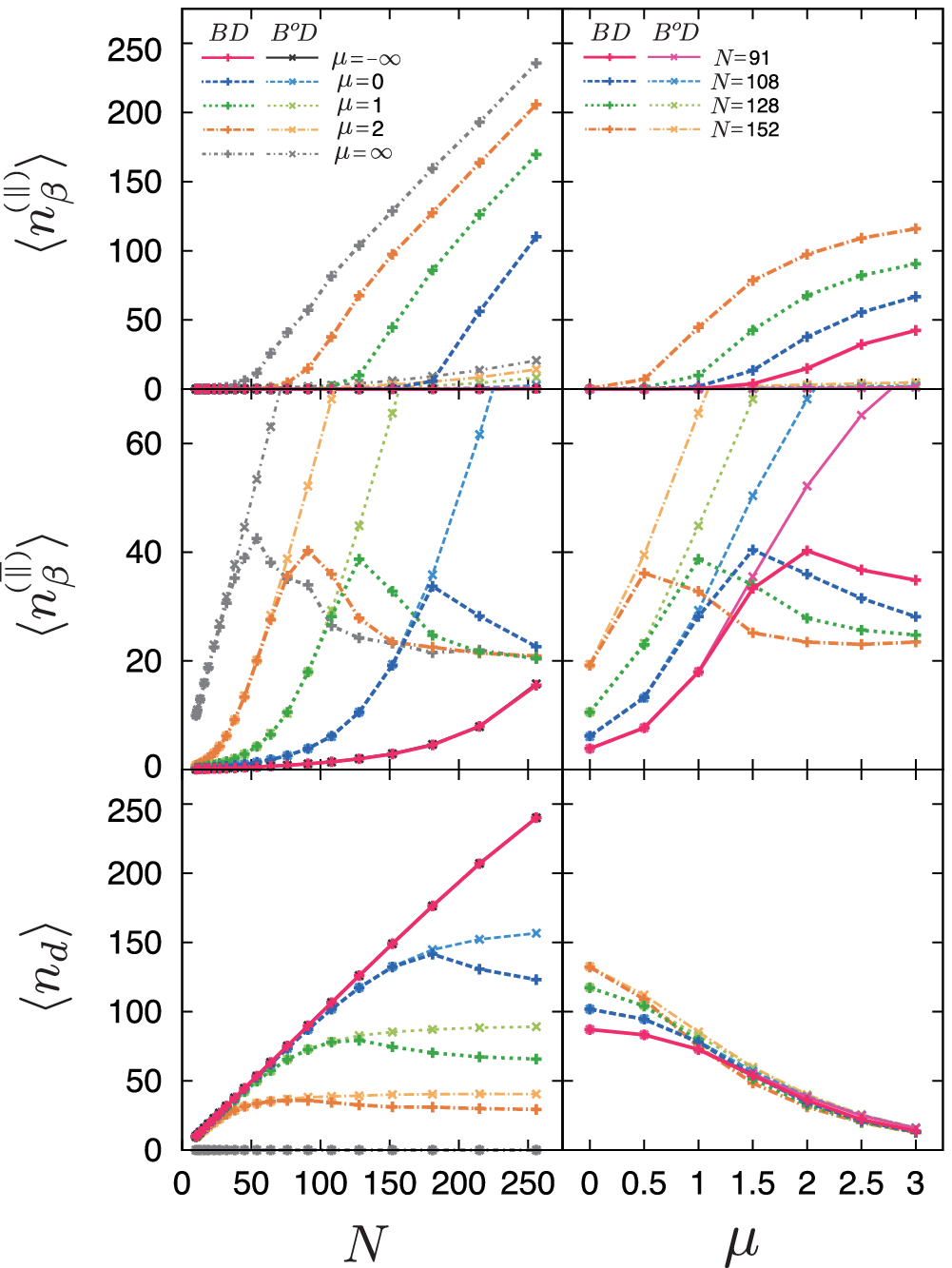}
}
\caption{
$N$ and $\mu$ dependences of $\alpha$-synuclein states for the $BD$ and $B^o \! D$ models with $L=64$. 
\label{fig:BD_ave_multiplot_L64}
}
\end{figure}

\renewcommand{\thefigure}{C3}
\begin{figure}[h]
\centerline{
\includegraphics[width=0.3\textwidth]{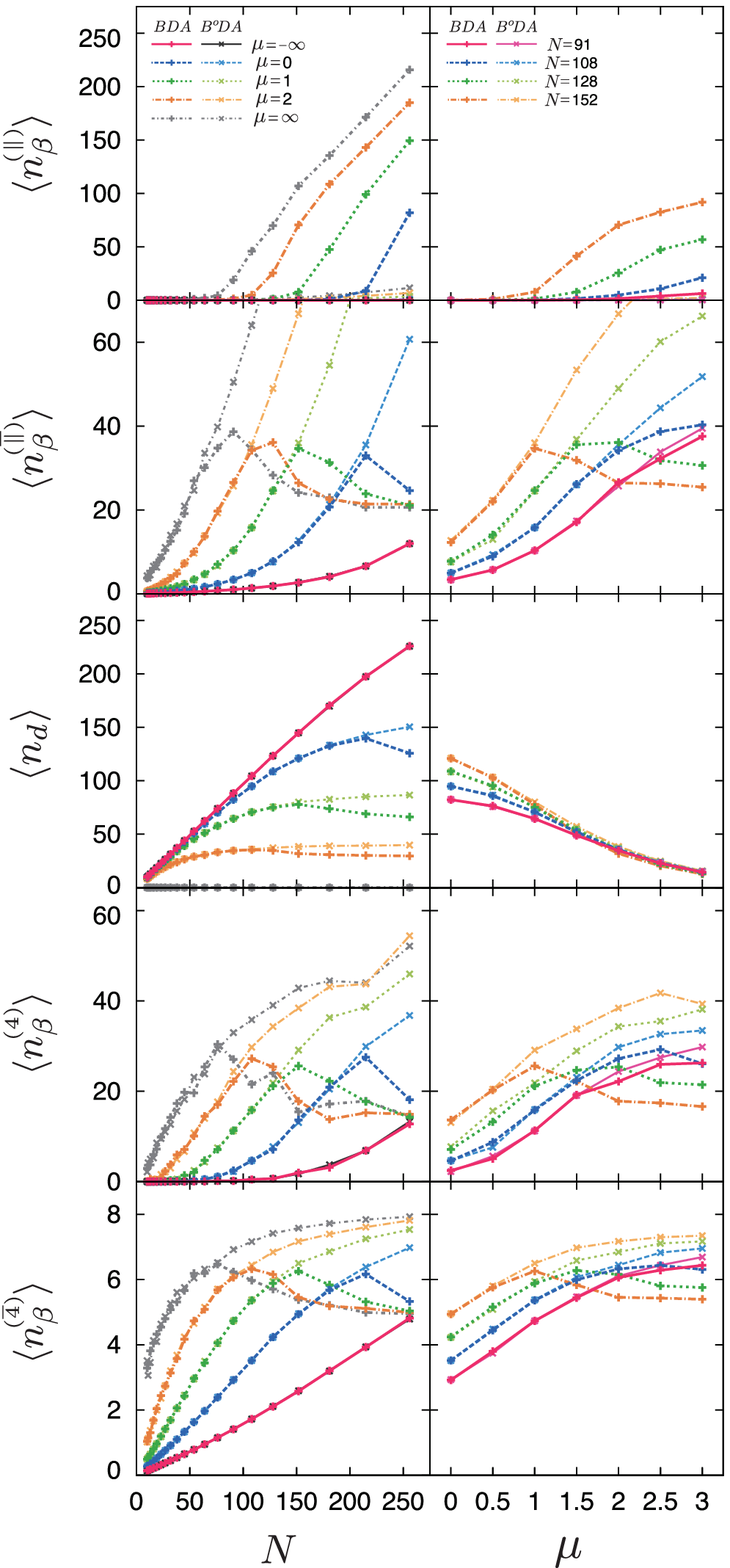}
}
\caption{
$N$ and $\mu$ dependences of $\alpha$-synuclein states for the $BDA$ and $B^o \! DA$ models with $L=64$. 
\label{fig:BDA_ave_multiplot_L64}
}
\end{figure}

\renewcommand{\thefigure}{C4}
\begin{figure}[h]
\centerline{
\includegraphics[width=0.3\textwidth]{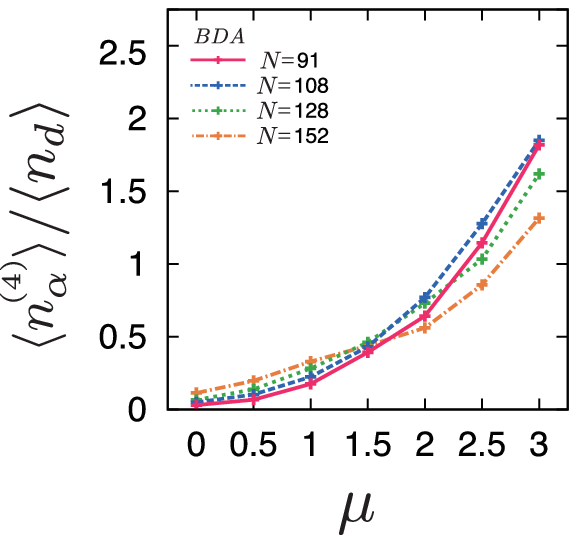}
}
\caption{
$N$ and $\mu$ dependences of the ratio between $\langle n^{(4)}_\alpha \rangle$ and $\langle n_d \rangle$ for the $BDA$ and $B^o \! DA$ models with $L=64$.
\label{fig:tetramer_disorder_ratio_L64}
}
\end{figure}

\end{document}